\documentclass[aps,prb,twocolumn]{revtex4}
\usepackage{amsmath}
\usepackage{graphicx}
\usepackage{bm}
\usepackage{enumerate}
\usepackage{hyperref}

\usepackage{floatrow}
\newfloatcommand{capbtabbox}{table}[][\FBwidth]

\begin{document}

\title{Finding hidden order in spin models with persistent homology}

\begin{abstract}
Persistent homology (PH) is a relatively new field in applied mathematics that studies the components and shapes of discrete data. In this work, we demonstrate that PH can be used as a universal framework to identify phases in spin models, including hidden order such as spin nematic ordering and spin liquids. By converting a small number of spin configurations to barcodes we obtain a descriptive picture of configuration space. Using dimensionality reduction to reduce the barcode space to color space leads to a visualization of the phase diagram.
\end{abstract}

\author{Bart Olsthoorn$^1$}
\email{bartol@kth.se}
\author{Johan Hellsvik$^1$}
\email{hellsvik@kth.se}
\author{Alexander V. Balatsky$^{1,2}$}
\affiliation{$^1$Nordita, KTH Royal Institute of Technology and Stockholm University, Roslagstullsbacken 23, SE-106 91 Stockholm, Sweden\\
$^2$Department of Physics, University of Connecticut, Storrs, CT 06269, USA}
\date{\today}
\maketitle

\section{\label{sec:intro}Introduction}

In condensed matter physics we often deal with materials where collective degrees of freedom such as magnetization, density modulations, coherent condensate formation, etc.  can undergo order to disorder transitions as a function of external parameters.
The distinctly different orderings of these degrees of freedom constitute different phases of the system, characterized by observables known as order parameters.

In the Landau theory of phase transitions, the transition between phases involves the breaking of a symmetry of the system. A classical example is the breaking of rotational symmetry in ferromagnetic materials. In the paramagnetic (high temperature) phase the spins are fully disordered and the system has the symmetry of the crystalline structure. At the onset of ferromagnetic order (low temperature) the symmetry is reduced. Order parameters reveal key symmetries of the phases and serve as indicators for the phase transitions where the condensed matter system undergoes a qualitative change as a result of change in control parameters such as temperature, pressure, and electromagnetic fields. Conventionally, these order parameters are constructed by hand.

In the presence of frustration between atomic spins, magnetic materials feature rich phase diagrams with different forms of collinear and noncollinear antiferromagnetic orderings, as well as more unconventional phases that lack long range magnetic orderings.  At the same time we know that both the easily identifiable phases, such as the ferromagnet, and more complicated and not easily visualized orders, like spin nematic, do describe strong correlations between spins. Of relevance for the present work are classical and quantum spin liquids \cite{2010itf,Balents2010,Savary2017,Taillefumier2017,Broholm2020}, the spin-ice phase \cite{Bramwell2001,Shannon2010,Edberg2019}, and nematic ordering that involves breaking of spin rotational symmetry while in the absence of magnetic ordering \cite{Chandra1990,Shannon2010,Capati2011,Voll2015,Kohama2019}. The spin nematic ordering is an example of a hidden order parameter which can be difficult to observe directly in an experiment, but that still reveals itself in the heat capacity signature and other observables at the  phase transition into nematic state \cite{Taillefumier2017,Aeppli2020}.

The commonly used indicators of phase transitions are heat capacity, susceptibility of the relevant order parameter(s), and derived quantities such as the reduced fourth-order Binder cumulant \cite{Binder1981}. While a peak in the heat capacity as a function of temperature can reveal the presence of a phase transition, it can not alone be used to characterize the phases. Understanding of the order parameter with concomitant susceptibility is a required step.

A new paradigm to investigate phase transitions based on neural networks recently emerged. It has been shown that neural networks can classify spin configurations sampled from a Monte Carlo simulation and thus can function as order parameter identifiers \cite{Carrasquilla2017,Iakovlev2019}. The neural network is trained in a supervised fashion, i.e. it is given labeled microstates, where the labels are determined by a known conventional order parameter.
Intriguingly, recent work  has shown that it is also possible to identify phases without prior information of the phase diagram, including phases without conventional order parameters \cite{Greitemann2019,vanNieuwenburg2017,Balabanov2020}.

In this work, we use a qualitatively different approach to {\em automatically construct} order parameters, based on a persistent homology (PH), a recently developed field in applied mathematics. In this approach the objective is to identify components and shapes given discrete data and a distance metric. These features are captured in a persistence diagram or barcode, two different graphical representations of the same information. Roughly speaking, these diagrams show at what scale different features are present in the data, and over what range of scale they persist.

PH is used in many fields including image processing \cite{10.1007/978-3-642-38221-5_19}, complex networks \cite{Horak_2009} and natural language processing \cite{Zhu_inproceedings}. However, it has not seen much use in condensed matter yet. In cosmology, it has been applied to study the cosmic web and its structure \cite{Sousbie2011}. In materials science, it is used to study nanoporous materials \cite{Lee2017} and other larger structures such as silica glass and polymers \cite{Buchet2018, pmlr-v48-kusano16}. A recent work proposes persistent homology observables to study dynamics of a quantum many-body system, using the two-dimensional Bose gas as an example \cite{spitz2020finding}.
There are two works that discuss the application of persistent homology for the detection phase transitions. Firstly, Donato \emph{et al.} \cite{Donato2016} study the mean-field XY model and a classical $\Phi^4$ model. The phase transitions are detected by computing the persistent homology of the configuration space with molecular dynamics simulations. Secondly, the recent work by Tran \emph{et al.} \cite{Tran2020} demonstrates the detection of phase transitions in the classical XY model and in quantum models.

We demonstrate how PH provides an unsupervised method for constructing phase diagrams, capturing both conventional and unconvential phases with a single framework. Each microstate is converted into a persistence diagram (or barcode), after which two different microstates can be compared with a a similarity metric on their corresponding persistence diagram.
Defining a similarity metric (or kernel) for persistence diagrams is still an open problem and we examine the capability of a sliced Wasserstein distance for spin models. This provides a unbiased way to construct a phase diagram.

The paper is organized as follows. We introduce persistent homology as an {\em order parameter} in Sec. \ref{sec:phorder}, describe our method to construct phase diagrams in Sec. \ref{sec:phaseconstr} and the XXZ model on pyrochlore lattice in Sec. \ref{sec:XXZpyro}. The results from the application of persistent homology to the XXZ model are presented in Sec. \ref{sec:phXXZ} followed by a discussion in Sec. \ref{sec:discussion} and conclusion in Sec. \ref{sec:conclusion}.

\section{\label{sec:phorder}Persistent homology as an order parameter}
Persistent homology (PH) is a method that identifies the qualitative features of a finite metric space (also called a point-cloud dataset) -- for an introduction see \cite{Otter2017,Ghrist2007}. Given a finite metric space and a distance parameter $r$, the point-cloud is converted into a graph where edges are added between vertices with distance smaller than $r$. A simplicial complex $K_1$ is built from the graph, and the homology of $K_1$ can be computed efficiently by linear algebra. Performing this construction for increasing $r$ yields a sequence of nested subcomplexes,
\begin{equation}
    K_1\subset K_2\subset \dots \subset K_l=K,
\end{equation}
where the sequence $K$ is called a filtered simplicial complex. The computed homology group $H_k(K_i)$ identifies $k$-dimensional holes in the simplicial complex. For example, $k=0$ relates to connected components (clusters) and $k=1$ to 1-dimensional holes and so on. The rank of $H_k$ counts the number of $k$-dimensional holes, also referred to as the Betti number $\beta_k$. The persistence (or lifetime) of $k$-dimensional holes over the distance parameter $r$ is visualized as a barcode, persistence diagram (PD), or lifetime diagram, all of which represent the same information.

Common types of complexes are the Vietoris-Rips (VR), \v{C}ech and Alpha complex \cite{Otter2017,Ghrist2007}. The complexes differ in number of simplices, affecting the computational complexity. We examine the capability of Alpha complex (also $\alpha$-complex) that is often used when the input data is 2D or 3D, in our case spins on a 3D lattice. The $\alpha$-complex is equivalent to the \v{C}ech complex when studying persistent homology, but it contains fewer simplices.

The $\alpha$-complex is constructed from a Delaunay triangulation on the input point-cloud dataset~\cite{gudhi:AlphaComplex}. For a 3-dimensional point cloud, the highest dimensional simplex is the 3-simplex (i.e. tetrahedron). Due to the input space being 3-dimensional, there are three homology groups of interest: $H_0, H_1$ and $H_2$.

With this brief introduction we now move to outline how we apply PH to spin models. The spin configurations (microstates) from classical Monte Carlo simulations form a metric space in multiple ways. Firstly, a single spin configuration can be considered a point in state space (e.g. discrete in the case of an Ising model, continuous in case of a Heisenberg model). Given a distance metric between two states, the corresponding persistence diagram (PD) is describing the topology of the phase diagram, where the homology groups describe the shapes of the phases in parameter space. Secondly, a single site in the spin system can be considered a point, leading to a single PD per microstate. A phase diagram can be constructed by comparing PDs.

In this work, we focus on the latter concept, i.e. the homology groups describe a microstate (spin configuration) directly. Averaging over microstates is possible by simply merging barcodes and discarding the the information from which microstate each bar (or point in PD) originated.

The concept introduced above requires a distance metric between two sites, where we can use the information of the lattice and the spin. The symmetry of the barcode is dictated by the choice of distance metric and determines its capability of capturing distinct phases.

As a simple example, consider the ferromagnetic square lattice Ising model ($J=1$) hosting a low-temperature symmetry-breaking ferromagnetic phase and a phase transition at $T_c\approx 2.269~J$ to a paramagnetic phase \cite{Onsager1944}. Choose a distance parameter $r$ where aligned spins are closer than opposite spins, with the distance being an Euclidean distance between spin tops. This  definition of a distance  metric leads to domains forming connected components in homology group $H_0$. Hence the Betti number $\beta_0(r)$ counts the number of domains present in the spin configuration, similar to the conventional order parameter of magnetization density $\left<M\right>$. In general, any metric capturing the shapes present in the spin texture is able to identify the two phases.

The phase diagram of a 2D Ising model discussed above is a convenient example, but the phase diagram is too simple to benefit from an approach with PH.
In this work we study a more complex system, namely the XXZ model on a pyrochlore lattice, known to host a large number of competing phases \cite{Taillefumier2017}.

\section{\label{sec:phaseconstr}Phase diagram construction}
\begin{figure}[b]
    \centering
    \includegraphics[width=1.0\linewidth]{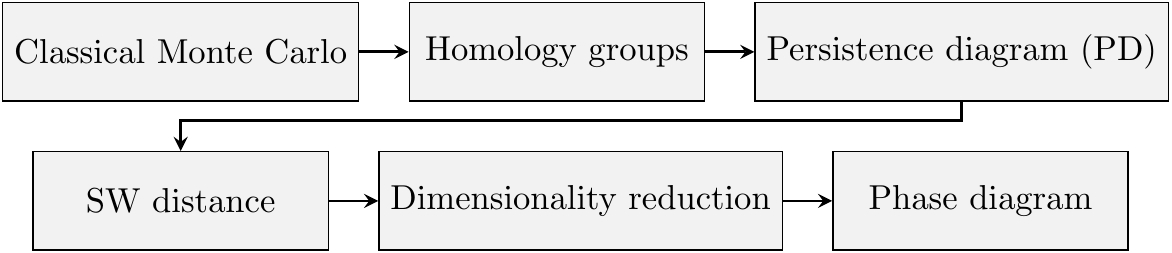}
    \caption{Summary of phase diagram construction.}
    \label{fig:scheme_summary}
\end{figure}

Traditionally, phase diagrams are constructed by observing a change in the order parameter that typically ranges from zero (phase A) to finite (phase B). In our proposed scheme, shown schematically in Fig. \ref{fig:scheme_summary}, the persistence diagram (PD) plays the role of the order parameter, and therefore we need to quantify the change in the PD.

Common distance metrics for PDs are the bottleneck distance and the Wasserstein distance \cite{Edelsbrunner2010cta}.
Both essentially measure the similarity by attempting to match the points (i.e. persistent features) in the two diagrams. These metrics are too computationally expensive for our purposes. Instead, we use the sliced Wasserstein (SW) distance which is an approximation of the Wasserstein distance \cite{Carriere2017}. The distance is measured between two PDs of same $k$th homology group, i.e. $\text{SW}(\text{PD}^k_i, \text{PD}^k_j)$. For the total distance, we consider holes ($H_1$) and voids ($H_2$):
\begin{equation}
    D_{ij} = \text{SW}(\text{PD}^1_i,\text{PD}^1_i) + \text{SW}(\text{PD}^2_i,\text{PD}^2_i)
    \label{eq:sw_distance_combined}
\end{equation}
The distance matrix $D$ is established by calculating the SW distance between all pairs of system parameters (e.g. all combinations of interaction $J$ and temperature $T$). Finally, the phase diagram is visualized by dimensionality reduction on $D$ to a 3-dimensional color space (red, green and blue). There are obvious drawbacks to this simple approach such as the non-linearity of color perception, but we find it sufficient. When distances are small and more color space is required, it is possible to limit the visualization to a subregion of the phase diagram, as shown in Sec.~\ref{sec:phXXZ}.

Dimensionality reduction algorithms aim to construct a low-dimensional image where distances of the original high-dimensional space are preserved as best as possible. Principal component analysis (PCA) is a common method based on matrix factorization, when given a set of input data points. In our case we only have a distance matrix $D$. Therefore PCA cannot be used directly, but multidimensional scaling (metric MDS) is the equivalent basic technique in this case \cite{Williams2002}. In summary, metric MDS algorithm is used to reduce the distance matrix $D$ to three RGB color channels that color-code the different phases in the phase diagram. Similar phases with small SW distances will appear close in color space and are therefore color-coded with similar colors.

\section{\label{sec:XXZpyro}XXZ model on the pyrochlore lattice}
The pyrochlore lattice is a cornerstone in research on frustrated magnetism, and is having a lead role in experimental and theoretical explorations of spin-ice physics \cite{Bramwell2001,2010itf,Shannon2010,Taillefumier2017,Edberg2019}. Motivated by the different chemistry of pyrochlore materials, a variety of spin Hamiltonians with long or short ranged interactions on the pyrochlore lattice have been investigated. The XXZ model on a pyrochlore lattice is a system with short ranged interactions having a rich phase diagram with competing antiferromagnetic, spin-ice, spin-liquid and spin-nematic phases ~\cite{Bramwell2001,Taillefumier2017}.
This model has also been used previously to test a machine learning model (support vector machine, SVM) that identifies phases~\cite{Greitemann2019}. The Hamiltonian of the XXZ model is given by
\begin{equation}
    H_{\text{XXZ}}=\sum_{\left<i,j\right>}J_{zz}S_{i,z}S_{j,z}-J_{\pm} \left(S_i^+S_j^- - S_i^-S_j^+\right),
    \label{eq:xxz}
\end{equation}
with $S_i^\pm = S_{i,x}\pm iS_{i,y}$ and $\bm{S}_i=\left(S_{i,x},S_{i,y},S_{i,z}\right)$, $\lVert\bm{S}_i\rVert=1$, and has the cubic symmetry of the pyrochlore lattice. In the following, energies and temperatures are expressed in terms of the antiferromagnetic $J_{zz}=1$ exchange interaction.

Extensive classical Monte Carlo simulations, field theoretical analysis, and spin dynamics simulations by Taillefumier \emph{et al.} \cite{Taillefumier2017} have established a $J_{\pm}-T$ phase diagram with six phases which here will be briefly recapitulated (numbering corresponding to \cite{Greitemann2019} and naming to \cite{Taillefumier2017}):

\begin{enumerate}[I]
    \item Easy-plane antiferromagnet (AF$\perp$). Spins lie in the plane perpendicular to the local $z$ axes. The phase occurs for large positive value of the ratio $J_\pm/J_{zz}$.
    \item Paramagnetic (PM). No long range spin ordering. Spins point in random directions with exponential decay of spin correlations.
    \item Easy-plane spin liquid (SL$\perp$) with algebraic spin-spin correlations, no long range magnetic ordering.
    \item Easy-plane spin-nematic (SN$\perp$) with algebraic spin-spin correlations, no long range spin dipole ordering.
    The rotational $U(1)$ symmetry of the local $z$ axes is broken by the onset of a higher order multipolar ordering.
    \item Spin ice (SI). Each tetrahedron on the lattice has the "two-in, two-out" spin configuration with the spins aligned along their local $z$ axis (see Fig.~\ref{fig:spin_ice}~(a)).
    The phase has algebraic spin correlations, no long range spin ordering.
    \item Pseudo-Heisenberg antiferromagnet (pHAF) with algebraic spin correlations distinct from phases III and V, no long range spin ordering.
\end{enumerate}

In summary, one phase has long range antiferromagnetic ordering (I AF$\perp$), one phase is disordered (II PM), one phase has spin-nematic ordering (IV SN$\perp$), and three of the phases (III SL$\perp$, V SI, and VI pHAF) are classical spin liquids. Relative to the paramagnetic phase, in which the system has the symmetries of the pyrochlore lattice, breaking of symmetry occurs only on entering the easy-plane antiferromagnetic phase (I AF$\perp$) or on entering the easy-plane spin-nematic phase (IV SN$\perp$), hence only when crossing in and or out from one of these two phases does the system go through a phase transition, with characteristic peak structure in the heat capacity and order parameter susceptibilities \cite{Taillefumier2017}. Transitions into or from one of the three spin liquid phases constitute a gradual evolution (crossover) of spin configurations, involving no breaking of symmetry, and lacks sharp peak structures in the heat capacity. The latter quantity nevertheless contains information that can be used to define a distinct criteria for where the crossover occurs \cite{Taillefumier2017}.
The three classical spin liquid phases can be distinguished by the different form of algebraic spin correlation reported for each of the phases \cite{Taillefumier2017}, correlations that can be measured in neutron scattering experiments. Rather remarkably, the easy-plane spin-nematic phase have the same algebraic spin correlation as the easy-plane spin liquid phase, yet the phases differ in the regard that in the latter the rotational symmetry in the local $z$ axes is broken by the nematic order parameter.

To test the proposal of PH for detection of the phases, we use representative spin configurations as an input. Spin configurations to construct the phase diagram were sampled within classical Monte Carlo simulations. In the Metropolis-Hastings algorithm, a spin flip is accepted if energy $E$ is lowered, otherwise accepted with probability $\exp{\left(-\beta\Delta E\right)}$ \cite{Sandvik2010}, where $\beta=1/T$ is the inverse temperature. The number of sites is given by $N=16 L^3$ with $L^3$ the number of cubic unit cells, each containing 16 sites, with each site having 6 nearest neighbors.
The simulations are carried out with $L=4$, i.e. $N=1024$ spins. Spin configurations are sampled on a parameter space grid $(T/J_{zz},J_\pm / J_{zz})$. Similar to \cite{Greitemann2019}, we sample 17 temperatures $T/J_{zz}$ with logarithmically spaced points between $10^{1}$ and $10^{-3}$. Exchange interactions $J_\pm / J_{zz}$ are sampled with 29 linearly spaced points $[-1, 0.4]$. Appendix~\ref{sec:impXXZ} lists the details of the implementation and the open-source code.

\section{\label{sec:phXXZ}Persistent homology on the XXZ model}

\begin{figure}[b]
    \centering
    \includegraphics[width=1.0\linewidth]{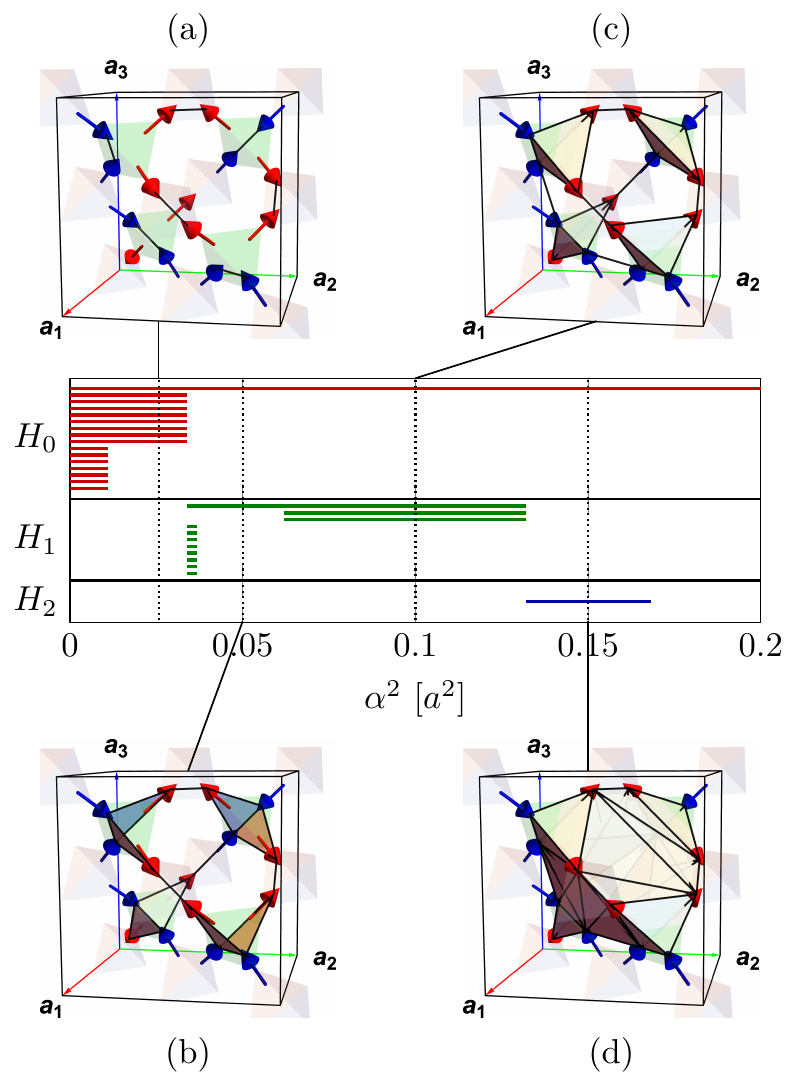}
    \caption{
    Barcode for a single unit cell with an exact spin ice configuration "two-in, two-out". The spins are all aligned to the local $z$ axis and there is no long range order. Shaded tetrahedrons form crystal lattice. The $\alpha$-complex at certain $\alpha^2$ filtration values is displayed. The characteristic features in $H_1$ and $H_2$ are captured by the Wasserstein distance to distinguish the spin ice phase from other phases.}
    \label{fig:spin_ice}
\end{figure}

A single spin configuration is converted into a point cloud by placing a point at each spin tip, where the spin length is set to 1/4 of the lattice tetrahedron edge length $d=\frac{a}{2\sqrt{2}}$. Euclidean distance is used once the point cloud is established using the distance metric
\begin{equation}\label{eq:distmet}
    \mathcal{D}(i,j)=r(i,j)+\frac{a}{2\sqrt{2}}\frac{\lVert \bm{S}_i-\bm{S}_j\rVert}{4},
\end{equation}
where $r(i,j)$ is the Euclidean distance between two sites. This corresponds to the distance of tips of the spin arrows (see Fig.~\ref{fig:spin_ice}). Note that $\bm{S}_i$ is in the global crystal frame, not in the local frame of Equation~\ref{eq:xxz} (see Appendix \ref{sec:impXXZ} for coordinate frame conversion). For nearest neighbors, this means $\mathcal{D}(i,j)$ is in the range $(d/2, 3d/2)$.

Figure~\ref{fig:spin_ice} shows the exact barcode for a single unit cell in the spin ice (V SI) phase, where spins are set according to the "two-in, two-out" rule. The connected components ($H_0$) reveal that there are two length scales present. The smaller length scale ($\alpha^2 \leq 0.010$) corresponds to neighboring spins forming 1-simplices (e.g. Fig.~\ref{fig:spin_ice}~(a)). The larger length scale ($\alpha^2\geq 0.034$) leads to a single connected component, as shown in Fig.~\ref{fig:spin_ice}~(b).
The 1-dimensional holes ($H_1$) appear at $\alpha^2 \geq 0.034$ as 9 bars. This includes 8 empty triangles (i.e. complex of 1-simplices rather than a single 2-simplices) and one large hole (also visible at $\alpha^2=0.05$, Fig.~\ref{fig:spin_ice}~(b)). The empty triangles are filled at $\alpha^2=0.037$ and become 2-simplices, shown as shaded triangles in Fig.~\ref{fig:spin_ice}~(b). At $\alpha^2=0.062$, the 1-simplices form a large tetrahedron with 2-simplices at its corners. A tetrahedron of 1-simplices has $\beta_1=\text{rank}(H_1)=3$, which corresponds to 3 bars for $H_1$ in Fig.~\ref{fig:spin_ice}~(c). All 1-dimensional holes are filled at $\alpha^2 \geq 0.131$. At this point, a single 2-dimensional void ($H_2$) appears for $\alpha^2=[0.132,0.168]$, shown in Fig.~\ref{fig:spin_ice}~(d).

Generally, each phase has persistent features at different characteristic length scales. We sample 32 statistically independent spin configurations for a single $(T/J_{zz},J_{\pm}/J_{zz})$ parameter combination. The $\alpha$-complexes are calculated by GUDHI \cite{gudhi:AlphaComplex}. The resulting barcodes are merged into a single barcode, discarding the information of the microstate origin from each bar. The similarity between barcodes is measured by the sliced Wasserstein (SW) distance, implemented in persim \cite{persim}, see Equation~\ref{eq:sw_distance_combined}.

\begin{figure}[t]
    \centering
    \includegraphics[width=\linewidth]{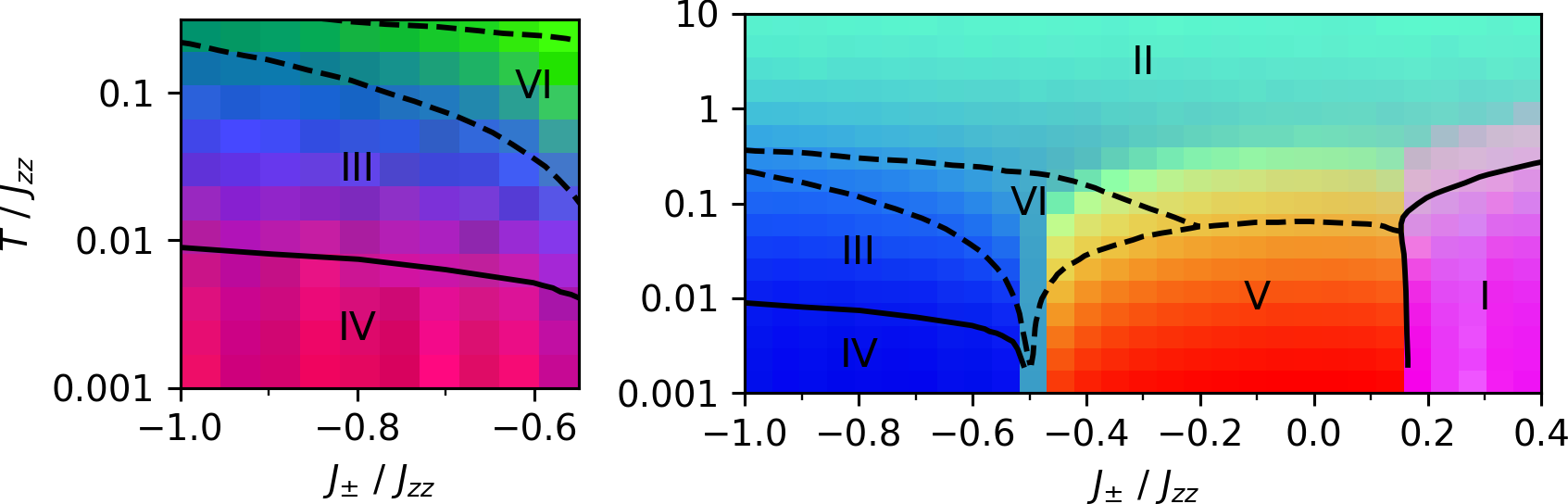}
    \caption{Phase diagram of the XXZ model on a pyrochlore lattice. Constructed by using dimensionality reduction (MDS) on the distance matrix of $(T/J_{zz},J_{\pm}/J_{zz})$ parameter space to the (R,G,B) color space. Sliced Wasserstein distance on homology groups is used. Crossover (transition) boundaries are drawn with solid (dashed) lines \cite{Taillefumier2017}.}
    \label{fig:ph_xxz_phase_diagram}
\end{figure}

The distance matrix $D$ is dimensionality reduced with MDS to a 3-dimensional space which is interpreted as red, green and blue channels. Figure~\ref{fig:ph_xxz_phase_diagram} shows the phase diagram. The paramagnetic phase II PM and the phases V SI and I AF$\perp$ are clearly distinct due to large pairwise sliced Wasserstein distances. The small distances between III SL$\perp$ and IV SN$\perp$ require a rescaling of the colormap to be visible, in the separate left pane. Also, the VI pHAF phase has small SW distances to the paramagnetic phase and is therefore difficult to detect, but a small color gradient is present.

\begin{figure}[b]
\centering
\includegraphics[width=1.0\linewidth]{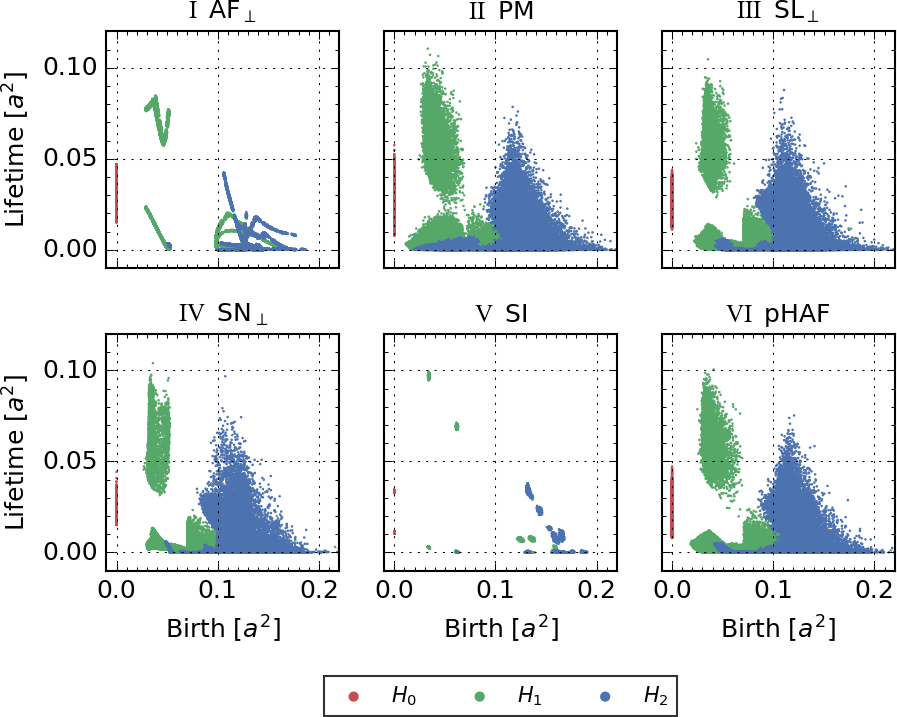}
\caption{Lifetime diagram of the $\alpha$-complexes constructed from spin configurations for different phases. The lifetime diagram shows the same information as a persistence diagram, with the diagonal along the x-axis, i.e. lifetime is death - birth. The corresponding ($J_{\pm}/J_{zz}$,$T/J_{zz}$) phase space parameters are $\text{I}=(0.4,0.001)$, $\text{II}=(0,10)$, $\text{III}=(-0.8,0.056)$, $\text{IV}=(-0.8,0.001)$, $\text{V}=(0,0.001)$ and $\text{VI}=(-0.5,0.001)$.}
\label{fig:lifetimes_phases_grid}
\end{figure}

Figure~\ref{fig:lifetimes_phases_grid} shows the lifetime diagrams and confirms that the phases II PM, III SL$_\perp$, IV SN$_\perp$ and VI SN$_\perp$ are similar, leading to smaller SW distances. A more detailed view of the lifetime diagrams as temperature decreases is shown in the Appendix Fig.~\ref{fig:full_lifetimes_grid}. Note that the spin ice phase V SI exhibits two $H_1$ features (with lifetimes around 0.07 and 0.1), corresponding to the two longer $H_1$ bar lengths in Fig.~\ref{fig:spin_ice}. In general, each phase has its own characteristic features and differences in its fingerprint are captured by the sliced Wasserstein distance. The qualitative similarities of phases and significant differences between lifetime diagrams, I AF vs V SI vs IV SN$_\perp$ vs III SL$_\perp$ vs VI pHAF phase, leads to the proposal that these lifetime diagrams can serve as respective order parameters (i.e. lifetime PH portraits of the phases).

\section{\label{sec:discussion}Discussion}
The PH method outlined here is using only 32 spin configurations for each point in parameter space $(T/J_{zz}, J_\pm / J_{zz})$ and does not require data on energies or heat capacities, making it an efficient technique. The $\alpha$-complexes are faster to calculate than the \v{C}ech and Vietoris-Rips complex and capture the holes and voids created by spins locally ordering. The computational bottleneck is the calculation of sliced Wasserstein distances between persistence diagrams. This could be improved by a faster distance metric for persistence diagrams, which is still an open problem in persistent homology. Finally, we note that persistent homology has very few free parameters in comparison to neural networks, which facilitates the interpretation of results.

In comparison to the work by Greitemann \emph{et al.} \cite{Greitemann2019} our method has a couple of key differences. Their work relies on the support vector machine (SVM) and the construction of monomials of spin components. We use persistent homology instead, involving no regression coefficients, and operate directly on the spin configurations. For the phase diagram construction, Greitemann \emph{et al.} use the Fiedler vector to partition the phase diagram given a distance matrix $D$. We instead use the simpler technique of multidimensional scaling (metric MDS) directly on the distance matrix $D$.

Recent work has shown that PH can detect phase transitions in a variety of classical and quantum models \cite{Donato2016,Tran2020}. We show the universality of the approach by identifying 6 different phases at once, including three classical spin liquids and construct a full two-dimensional phase diagram.

The detection of phases is possible by studying persistence diagrams (Fig.~\ref{fig:lifetimes_phases_grid}) and changes are quantified by the sliced Wasserstein distance. We reduce the persistence diagrams space to color space with matrix factorization, which proves to work best when SW distances are large and of similar magnitude. Alternatively, dimensionality reduction based on the neighbor graph approach (e.g. Uniform Manifold Approximation and Projection, UMAP \cite{mcinnes2018umap-software}) could highlight local structure better at the expense of representing global structure of the original high dimensional barcode space.

\section{\label{sec:conclusion}Conclusion}
We demonstrate that persistent homology (PH) can be used to capture different types of order in spin models.
The application of our method to the XXZ model on the pyrochlore lattice with its six phases demonstrates the versatility of this approach.
Both the phase with long range order and the phases with only local ordering are identified.

The barcodes reveal the characteristic length scales present in the spin model. Using sliced Wasserstein distance, we obtain a distance matrix for all pairwise system parameters. The distance matrix can be visualized using dimensionality reduction, revealing similar regions in parameter space. Alternatively, the persistence diagrams can be inspected directly to observe changes as the system temperature decreases.

In summary, persistent homology provides a new general computational framework to study both long range and local spin ordering. Extending the persistent homology framework to quantum spin models will be the topic of future work.

\section{Acknowledgements}\label{acknowledgements}
The authors are grateful to M. Geilhufe, Jens H. Bardarson, and Q. Yang for discussions.
We acknowledge funding from the VILLUM FONDEN via the Centre of Excellence for Dirac Materials (Grant No. 11744), the European Research Council ERC HERO grant, University of Connecticut, and the Swedish Research Council (VR) through a neutron project grant (BIFROST, Dnr. 2016-06955).
The authors acknowledge computational resources from the Swedish National Infrastructure for Computing (SNIC) at the National Supercomputer Centre at Link\"oping University, the Centre for High Performance Computing (PDC), the High Performance Computing Centre North (HPC2N), and the Uppsala Multidisciplinary Centre for Advanced Computational Science (UPPMAX).

\bibliography{references}
\onecolumngrid
\appendix
\section{\label{sec:impXXZ}XXZ model on pyrochlore lattice implementation}

Monte Carlo requires calculating the change in energy when changing spin. The energy change when changing spin $S_i\rightarrow S'_i$ for the XXZ model (see Equation~\ref{eq:xxz}) on the pyrochlore lattice is
\begin{equation}
    \Delta H = \frac{2J_\pm}{J_{zz}}\left[\left(S_{x,i}-S'_{x,i}\right)\sum_{j}^6 S_{j,x}+
    \left(S_{y,i}-S'_{y,i}\right)\sum_{j}^6 S_{j,y}\right]-\left(S_{z,i}-S'_{z,i}\right)\sum_{j}^6 S_{j,z}.
\end{equation}
Flipping $N$ random spins constitutes a Monte Carlo (MC) step, where $N=L^3$ is the total number of sites.

The parameter space $(T/J_{zz}, J_\pm / J_{zz})$ is sampled in a similar manner as \cite{Greitemann2019}. For each exchange interaction value, we start by initiating 32 random (paramagnetic) high temperature configurations. After each temperature reduction, $10^5$ MC steps are performed and a single spin configuration is sampled. Therefore, for each parameter combination, we sample 32 uncorrelated spin configurations, i.e. each spin configuration originates from an independent random initialization.

The spins in Hamiltonian (Equation~\ref{eq:xxz}) are defined in a local coordinate frame, which can be converted into a global frame,
\begin{equation}
    \bm{S}_i=\bm{x}_i^\text{local}S_{x,i}+\bm{y}_i^\text{local}S_{y,i}+\bm{z}_i^\text{local}S_{z,i}.
\end{equation}
All analysis in this work is done with spins in the global frame.

We implement the Monte Carlo simulation in the Julia programming language. The spin configurations are stored in the HDF5 format for subsequent analysis. The persistent homology calculations are performed in Python, using the GUDHI package for alpha complexes and persim for the sliced Wasserstein distance. All the code is available at \url{https://github.com/bartolsthoorn/PH_XXZ}.

\begin{figure}[h]
\begin{floatrow}
\ffigbox{%
  \includegraphics[width=0.5\linewidth]{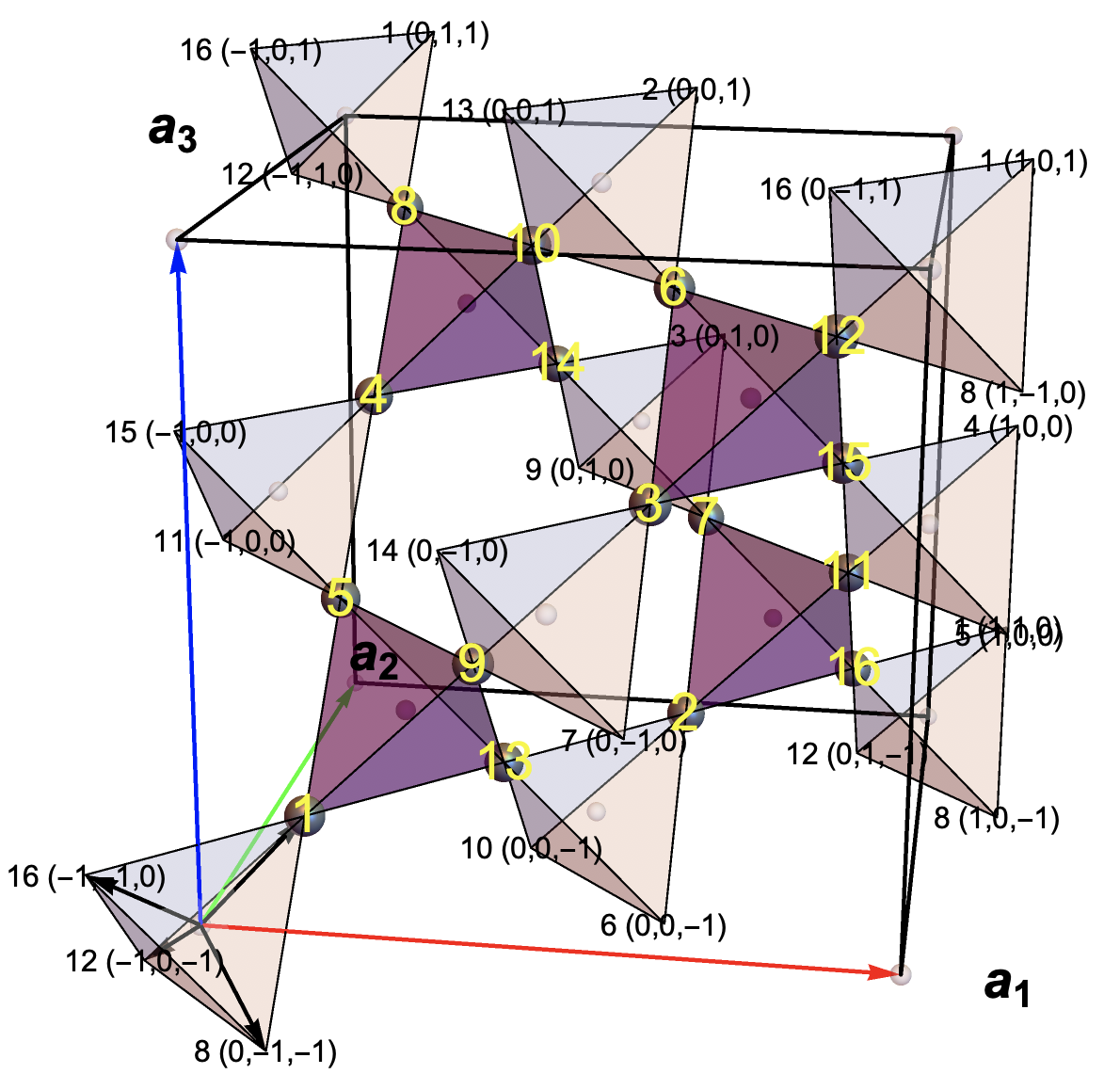}%
}{%
  \caption{Pyrochlore lattice. Each site has 6 nearest neighbors. The three values in parenthesis label the unit cell offset of the neighboring site.}%
}
\capbtabbox{%
  \begin{tabular}{c|c|c||c|c|c}
    ID & Position $\frac{a}{8}\cdot$ & $S_i$ & ID & Position $\frac{a}{8}\cdot$ & $S_i$\\
    \hline
1 & (1,1,1) & 0 & 9 & (3,1,3) & 2\\
2 & (5,5,1) & 0 & 10 & (3,5,7) & 2 \\
3 & (5,1,5) & 0 & 11 & (7,5,3) & 2\\
4 & (1,5,5) & 0 & 12 & (7,1,7) & 2\\
5 & (1,3,3) & 1 & 13 & (3,3,1) & 3\\
6 & (5,3,7) & 1 & 14 & (3,7,5) & 3\\
7 & (5,7,3) & 1 & 15 & (7,3,5) & 3\\
8 & (1,7,7) & 1 & 16 & (7,7,1) & 3
    \end{tabular}
}{%
  \caption{Positions of the pyrochlore lattice.}%
}
\end{floatrow}
\end{figure}

\section{\label{sec:lifetime_phase_diagram}Lifetime diagrams in phase diagram}
\begin{figure}[h]
\centering
\includegraphics[width=0.55\linewidth]{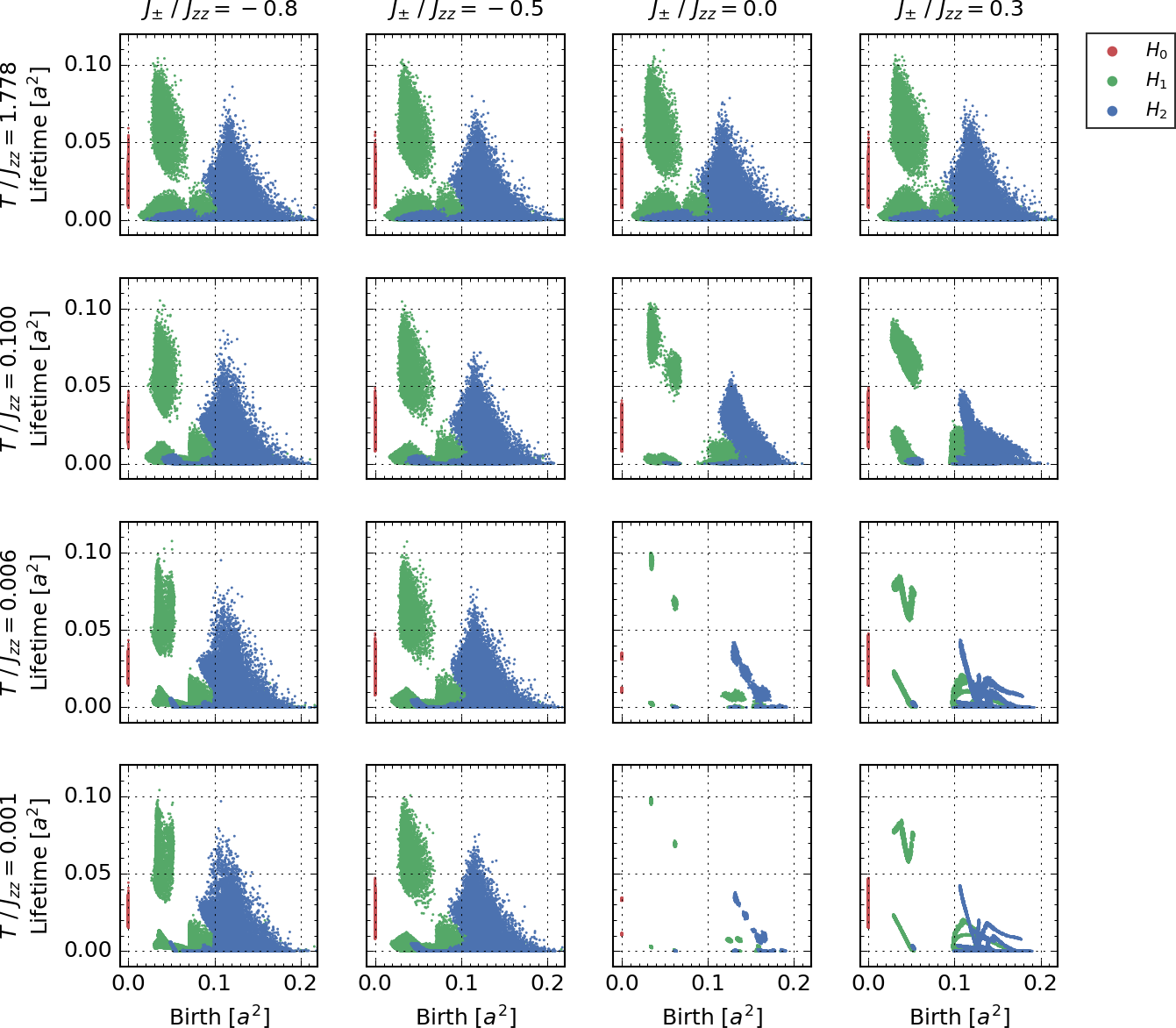}
\caption{Lifetime diagram of the $\alpha$-complexes constructed from spin configurations at different phase space parameters. At the lowest temperature $T/J_{zz}=0.001$ (bottom row), persistent features at characteristic length are visible.}
\label{fig:full_lifetimes_grid}
\end{figure}

Figure~\ref{fig:full_lifetimes_grid} shows how the lifetime diagram changes as temperature is decreased. The selected temperature and exchange parameter are listed at the beginning of each row and column, respectively. Note that a single lifetime diagram includes 32 sampled spin configurations. For each spin configuration a barcode is produced, and each bar is treated the same, leading to a single persistence diagram. In other words, diagrams are aggregated by dropping the information of the origin of each persistent feature.

\end{document}